\documentclass{aa}
\usepackage{graphics,epsfig,txfonts}

\usepackage{natbib}
\bibpunct{(}{)}{;}{a}{}{,}

\newcommand{\ergs}[1]{$\times 10^{#1}$ erg s$^{-1}$}
\newcommand{\oergs}[1]{$10^{#1}$ erg s$^{-1}$}
\newcommand{\hcm}[1]{$\times 10^{#1}$ cm$^{-2}$}

\newcommand{\expo}[1]{$\times 10^{#1}$}
\newcommand{\oexpo}[1]{$10^{#1}$}

\newcommand{\nh}{N$_{\rm H}$}

\newcommand{\ltsima}{$\buildrel < \over \sim$}
\newcommand{\lsim}{\lower.5ex\hbox{\ltsima}}
\newcommand{\gtsima}{$\buildrel > \over \sim$}
\newcommand{\gsim}{\lower.5ex\hbox{\gtsima}}
\newcommand{\xmm}{XMM-Newton}
\newcommand{\xtej}{\hbox{XTE\,J0103-728}}
\begin{document}
 
\title{XMM-Newton observations of the Small Magellanic Cloud:\\ 
       X-ray outburst of the 6.85 s pulsar XTE\,J0103-728
       \thanks{Based on observations with 
               XMM-Newton, an ESA Science Mission with instruments and contributions 
               directly funded by ESA Member states and the USA (NASA)}}
\author{F.~Haberl \and W.~Pietsch}

\titlerunning{XMM-Newton observations of XTE\,J0103-7283}
\authorrunning{Haberl \& Pietsch}
 
\institute{Max-Planck-Institut f\"ur extraterrestrische Physik,
           Giessenbachstra{\ss}e, 85748 Garching, Germany\\
	   \email{fwh@mpe.mpg.de, wnp@mpe.mpg.de}}
 
\date{Received / Accepted }
 
\abstract{A bright X-ray transient was seen during an XMM-Newton observation in the 
          direction of the Small Magellanic Cloud (SMC) in October 2006.}
         {The EPIC data allow us to accurately locate the source and to investigate 
	  its temporal and spectral behaviour.}
	 {X-ray spectra covering 0.2$-$10 keV and pulse profiles in different energy bands
	  were extracted from the EPIC data.}
         {The detection of 6.85 s pulsations in the EPIC-PN data unambiguously identifies
	  the transient with \xtej, discovered as 6.85 s pulsar by RXTE.
          The X-ray light curve during the XMM-Newton observation shows flaring activity 
	  of the source with intensity changes by a factor of two within 10 minutes.
	  Modelling of pulse-phase averaged spectra with a simple absorbed power-law indicates
	  systematic residuals which can be accounted for by a second emission component.
	  For models implying blackbody emission, thermal plasma emission or emission from 
	  the accretion disk (disk-blackbody), the latter yields physically sensible parameters.
	  The photon index of the power-law of $\sim$0.4 indicates a relatively hard spectrum. 
	  The 0.2$-$10 keV luminosity was 2\ergs{37} with a contribution of $\sim$3\% from
	  the disk-blackbody component. A likely origin for the excess emission is reprocessing
	  of hard X-rays from the neutron star by optically thick material near the inner edge of
	  an accretion disk.
          From a timing analysis we determine the pulse period to 6.85401(1) s indicating an 
	  average spin-down of $\sim$0.0017 s per year since the discovery of \xtej\ 
	  in May 2003.}
         {The X-ray properties and the identification with a Be star confirm \xtej\ as Be/X-ray 
	  binary transient in the SMC.}

\keywords{galaxies: individual: Small Magellanic Cloud --
          galaxies: stellar content --
          stars: emission-line, Be -- 
          stars: neutron --
          X-rays: binaries}
 
\maketitle
 
\section{Introduction}

The transient pulsar \xtej\ with a period of 6.8482$\pm$0.0007 s was discovered during 
RXTE observations of the Small Magellanic Cloud (SMC). 
The pulsations were detected on 2003, Apr. 29, May 7, May 15 and May 19 but not on Apr. 24
and May 28 \citep{2003ATel..163....1C}, suggesting an X-ray outburst which lasted about
three to four weeks. This behaviour is characteristic for Be/X-ray binaries undergoing 
a type II outburst which is caused by the ejection of matter from the Be star 
and enhanced accretion onto a compact object, in most cases a neutron star 
\citep{1998A&A...338..505N}. Shorter outbursts (type I) with a duration of a few days 
are often separated by the orbital period of the binary system. They generally occur close 
to the time of periastron passage of the neutron star when the neutron star approaches 
the circumstellar disc of the Be star \citep[see e.g.][]{2001A&A...377..161O}.

Be/X-ray binaries form the major class of High Mass X-ray Binaries (HMXBs). In the remaining 
systems - the supergiant HMXBs - the compact object accretes matter from the fast stellar wind 
of an early type O or B supergiant star. In the SMC more than 60 Be-HMXBs are known, while 
only one HMXB is established as supergiant system (SMC\,X-1). Recent reviews of the optical and X-ray properties
of these systems can be found in \citet{2005MNRAS.356..502C} and \citet{2004A&A...414..667H}, 
respectively.

The X-ray transient \xtej\ is one of several Be/X-ray binary candidates in the SMC 
which were discovered through their pulsations with RXTE. This non-imaging instrument 
is able to monitor the activity of pulsars in the SMC \citep{2005ApJS..161...96L}
but provides only a coarse determination of their sky coordinates. The latter hampers
the optical identification and final confirmation of the HMXB nature of the pulsars. The pulsar 
\xtej\ was detected in X-ray outburst during an \xmm\ observation of RX\,J0103.8$-$7254, 
a candidate super-soft X-ray source in the direction of the SMC discovered by ROSAT.
The precise localization in the EPIC images allowed us to identify the optical counterpart 
of \xtej\ which shows optical brightness and colours consistent with a Be star 
\citep[see][]{2007ATel.1095....1H}.
Analysis of MACHO and OGLE data of the counterpart reveals long-term variations in the 
optical light curve \citep{2007ATel.1181....1S} on time scales of 620 to 660 days which are 
not periodic, suggesting that they are related to the Be phenomenon and not to the binary orbit
\citep{2007arXiv0711.4010M}.

Here we present a temporal and spectral analysis of the EPIC data of \xtej\ obtained in 
October 2006. 

\section{Observations and data reduction}
\label{sect-obs}

%
%
%

\begin{table}
\caption[]{\xmm\ EPIC observations of \xtej.}
\begin{tabular}{cccc}
\hline\hline\noalign{\smallskip}
\multicolumn{1}{c}{Observation} &
\multicolumn{2}{c}{Pointing direction} &
\multicolumn{1}{c}{Sat.} \\

\multicolumn{1}{c}{ID} &
\multicolumn{1}{c}{R.A.} &
\multicolumn{1}{c}{Dec.} &
\multicolumn{1}{c}{Rev.} \\

\multicolumn{1}{c}{} &
\multicolumn{2}{c}{(J2000.0)} &
\multicolumn{1}{c}{} \\
\noalign{\smallskip}\hline\noalign{\smallskip}
 0402000101 & 01 03 52.2 & -72 54 28 & 1248  \\
\noalign{\smallskip}\hline\noalign{\smallskip}
\multicolumn{1}{c}{EPIC$^1$} &
\multicolumn{1}{c}{Start time} &
\multicolumn{1}{c}{End time} &
\multicolumn{1}{c}{Net} \\

\multicolumn{1}{c}{instrument} &
\multicolumn{2}{c}{2006-10-03 (UT)} &
\multicolumn{1}{c}{exposure} \\

\multicolumn{1}{c}{configuration} &
\multicolumn{1}{c}{} &
\multicolumn{1}{c}{} &
\multicolumn{1}{c}{ks} \\
\noalign{\smallskip}\hline\noalign{\smallskip}
 PN FF thin & 00:31:21 & 06:13:06 & 17.72 \\
 M1 FF thin & 00:08:39 & 06:12:46 & 21.32 \\
 M2 FF thin & 00:08:39 & 06:12:51 & 21.30 \\
\noalign{\smallskip}\hline\noalign{\smallskip}
\end{tabular}

$^1$ FF thin: full frame CCD readout mode with 73 ms frame time for PN and 2.6 s for MOS; thin optical blocking filter.
\label{tab-obs}
\end{table}

An X-ray transient was seen as brightest source in the EPIC field of view of an \xmm\ 
\citep{2001A&A...365L...1J} observation in the direction of the SMC, as
summarized in Table~\ref{tab-obs}. The EPIC-MOS \citep{2001A&A...365L..27T} and EPIC-PN 
\citep{2001A&A...365L..18S} cameras were operated in imaging mode 
covering the source at an off-axis angle of about 12\arcmin.
Background flaring activity was negligible and we used the full exposure time for 
our analysis.
For the data processing we used the \xmm\ Science Analysis System (SAS) version 7.1.0 
supported by tools from the FTOOL package together with XSPEC version 11.3.2p for spectral modelling.  

\section{Results}

We performed a source detection analysis of the EPIC images using standard maximum likelihood
procedures from the SAS package. After applying a bore-sight correction using a background
AGN and a foreground star 
in the field of view the position of the brightest source was determined to
R.A. = 01$^{\rm h}$02$^{\rm m}$53\fs39 and Dec. = --72\degr44\arcmin34\farcs6 (J2000.0) with
a remaining systematic uncertainty of 1.1\arcsec\ (1$\sigma$).
No source at this position was seen in ROSAT observations \citep{2000A&AS..142...41H,2000A&AS..147...75S}.

\begin{table*}
\caption[]{Optical identification of \xtej.}
\begin{center}
\begin{tabular}{lcrrrrr}
\hline\hline\noalign{\smallskip}
\multicolumn{1}{c}{Catalogue} &
\multicolumn{1}{c}{R.A. and Dec. (J2000.0)} &
\multicolumn{1}{c}{Vmag} &
\multicolumn{1}{c}{B$-$V} &
\multicolumn{1}{c}{U$-$B} &
\multicolumn{1}{c}{V$-$R} &
\multicolumn{1}{c}{V$-$I} \\

\noalign{\smallskip}\hline\noalign{\smallskip}
 UBVR & 01$^{\rm h}$02$^{\rm m}$53\fs30 --72\degr44\arcmin34\farcs9 & 14.59 & $-$0.08 & $-$0.96 & $-$0.90 & --      \\
 MCPS & 01$^{\rm h}$02$^{\rm m}$53\fs39 --72\degr44\arcmin34\farcs7 & 14.99 & $-$0.19 & $-$1.12 & --      & $-$0.10 \\
 OGLE & 01$^{\rm h}$02$^{\rm m}$53\fs29 --72\degr44\arcmin34\farcs8 & 14.68 & $-$0.12 & --	& --	  & 0.04 \\
\noalign{\smallskip}\hline
\end{tabular}
\end{center}
\label{tab-ids}
\end{table*}

\begin{figure}
  \resizebox{0.98\hsize}{!}{\includegraphics[clip=]{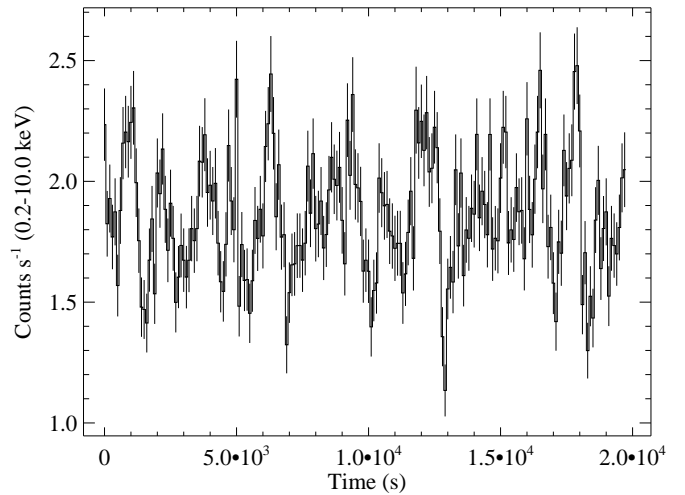}}
  \caption{Broad-band EPIC-PN light curve of \xtej. The data is background subtracted and binned to 100 s 
  with time 0 corresponding to MJD 54011.02678.}
  \label{fig-lcurve}
\end{figure}

During the \xmm\ observation the source showed strong flaring activity with intensity changes by a 
factor of $\sim2$ within 10 minutes. This is illustrated in Fig.~\ref{fig-lcurve} where the
broad band EPIC-PN light curve is shown. 
A Fourier analysis of the EPIC-PN data clearly revealed the presence of 
6.85 s pulsations in the X-ray flux which identifies the bright EPIC source
with \xtej. No significant signal of higher harmonics is seen.
Using a folding technique the pulse period is determined to (6.85401 $\pm$ 1\expo{-5}) s 
(1$\sigma$ error). This corresponds to an average spin-up of $\sim$0.0017 s per year
between the RXTE and the \xmm\ observations, 3.4 years apart.

The precise position allowed us to identify the optical counterpart 
\citep{2007ATel.1095....1H} which is included in the 
UBVR CCD Survey of the Magellanic Clouds \citep{2002ApJS..141...81M}, 
the Magellanic Clouds Photometric Survey \citep[MCPS,][]{2002AJ....123..855Z} and in
the OGLE BVI photometry catalogue \citep{1998AcA....48..147U} as summarized in Table~\ref{tab-ids}.
Optical brightness, colors and temporal properties are all consistent with a 
Be star \citep{2007ATel.1181....1S,2007arXiv0711.4010M}, confirming the Be/X-ray 
binary nature of \xtej.

\begin{figure}
  \resizebox{0.98\hsize}{!}{\includegraphics[clip=]{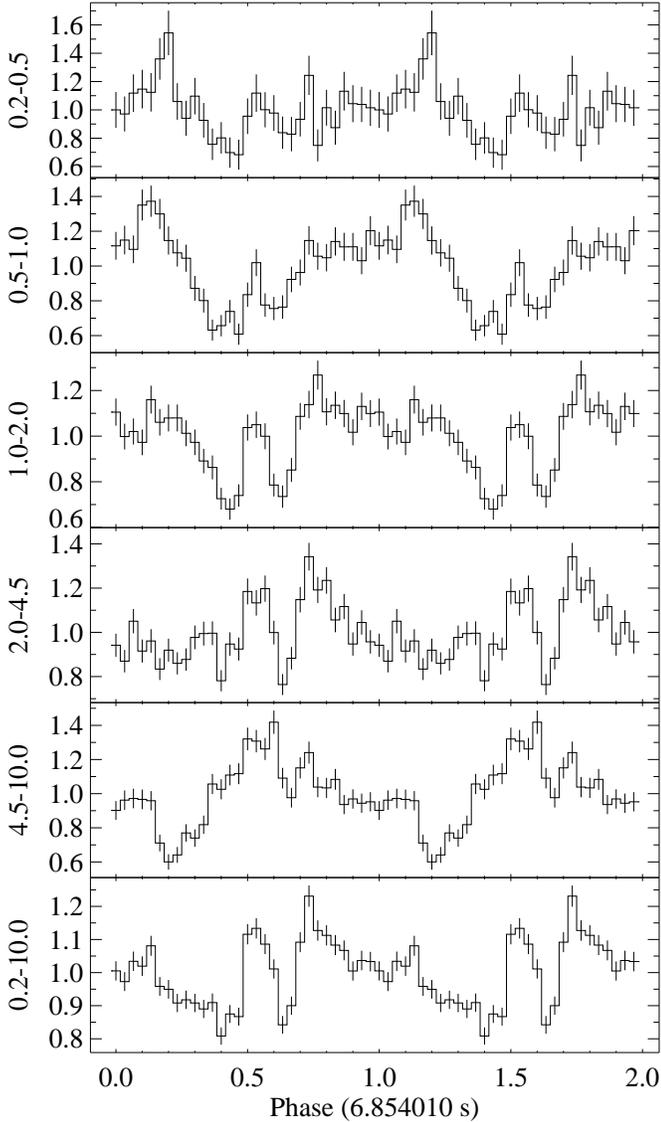}}
  \caption{Folded EPIC-PN light curves in the standard EPIC energy bands. The panels
           show the pulse profiles for the different energies specified in keV. The intensity profiles 
           are background subtracted and normalized to the average count rate 
	   (in cts s$^{-1}$: 0.103, 0.268, 0.482, 0.530, 0.466, 1.86 from top to bottom).}
  \label{fig-pulse}
\end{figure}

\begin{figure}
  \resizebox{0.98\hsize}{!}{\includegraphics[clip=]{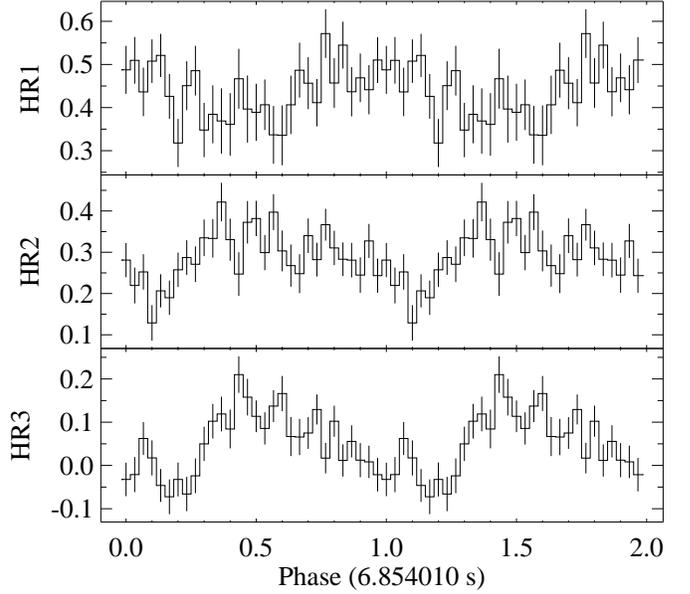}}
  \caption{Hardness ratios HR1, HR2 and HR3 derived from the folded EPIC-PN light curves shown 
           in Fig.~\ref{fig-pulse}.}
  \label{fig-hr}
\end{figure}

The X-ray light curves were folded in standard EPIC energy bands as shown in Fig.~\ref{fig-pulse}.
Hardness ratios were derived with HR1 = (R2-R1)/(R2+R1), HR2 = (R3-R2)/(R3+R2) 
and HR3 = (R4-R3)/(R4+R3) with RN denoting the background-subtracted count rate in band N
(starting with band 1 at the lowest energies). HR1, which is computed from count rates
below 1 keV is more sensitive to changes in absorption column density while HR2 and HR3 can be used as 
indicator for variations in the shape of the intrinsic source spectrum up to 4.5 keV.
The three hardness ratios plotted versus pulse phase are shown in Fig.~\ref{fig-hr}.

The pulse profile of \xtej\ is highly structured and strongly energy dependent. The pulsed fraction decreases 
with increasing energy up to $\sim$4.5 keV, but is higher again above $\sim$4.5 keV.
Several features in the folded light curves strongly change their appearance in the various 
energy bands. E.g. the intensity maximum in the 0.2$-$0.5 keV band reverses to a minimum at 
high energies above 4.5 keV and the sharp minima seen between 1.0$-$2.0 keV are largely smeared 
out at energies above 4.5 keV. HR1 indicates a relatively smooth change in the low-energy
part of the spectrum, either due to variations in column density or in a low-energy spectral component.
Variations are also seen in HR2 and similarly in HR3 which suggest changes in the overall spectral shape.


\begin{figure}
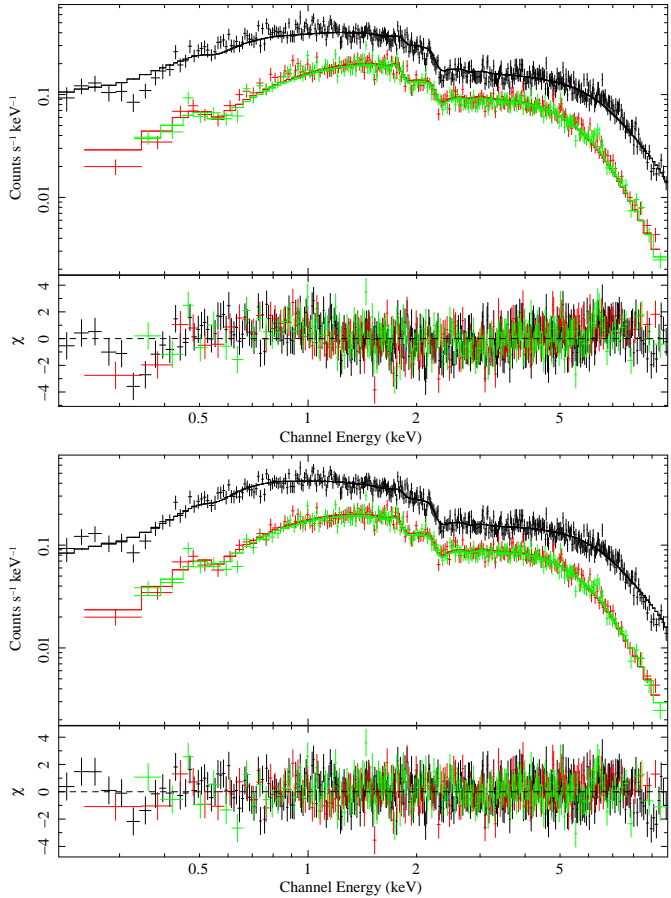

  \resizebox{0.98\hsize}{!}{\includegraphics[angle=-90,clip=]{spectrum_1_pow.ps}}
  \resizebox{0.98\hsize}{!}{\includegraphics[angle=-90,clip=]{spectrum_1_diskbb_pow.ps}}
  \caption{EPIC spectra of \xtej. EPIC-PN is shown in black and 
  EPIC-MOS in red (M1) and green (M2) (both grey in black and white representation). The histograms show
  the best-fit model: {\it (top)} an absorbed power-law and {\it (bottom)} an absorbed two-component model with 
  multi-temperature disk and power-law emission (PL+DB in Table~\ref{tab-spectra}).}
  \label{fig-spectra}
\end{figure}

We extracted pulse-phase averaged EPIC spectra for PN (single + double pixel events, PATTERN 0$-$4) and 
MOS (PATTERN 0$-$12) excluding bad CCD pixels and columns (FLAG 0). 
Due to the large off-axis angle we used an elliptical source extraction 
region which was cut in the case of EPIC-PN by the nearby CCD border. The fluxes derived from 
the PN spectrum were systematically lower than those from the MOS, probably due to under-correction
of the flux losses. Therefore, we report fluxes and luminosities (see below) derived from the MOS spectra.
The three EPIC spectra were simultaneously fit with the same model allowing for a 
constant normalization factor between the spectra. The spectra show an extremely flat distribution in the
EPIC energy band which indicates a very hard spectrum. We included two absorption components in
our spectral modelling, accounting for  the Galactic foreground absorption (with a fixed hydrogen column 
density of 6\hcm{20} and with elemental abundances from \citet{2000ApJ...542..914W}) and the SMC absorption 
(with column density as free parameter in the fit and with metal abundances reduced to 0.2 as 
typical for the SMC \citep{1992ApJ...384..508R}).
A bremsstrahlung model is not able to reproduce the flat spectral distribution while a
power-law (PL) yields a formally acceptable fit (see Table~\ref{tab-spectra} and Fig.~\ref{fig-spectra}).
However, the residuals indicate some systematic deviation from a power-law model.
Using a broken power-law yields a good fit with $\chi^2_{\rm r}$=1.02 with a steeper power-law 
index below 2 keV than above, which indicates excess emission below $\sim$2 keV. For a better description 
of this excess we add different model components to the simple power-law.

Two different kinds of low-energy excess seen in the X-ray spectra of HMXBs were reported in the 
literature. 
A soft emission component from Be-HMXBs is interpreted as re-processed X-ray emission
and may originate from different places in the binary system \citep{2004ApJ...614..881H}. 
The Be-HMXB RX\,J0103.6$-$7201 (with 1323 s the pulsar with the longest period known in the SMC) showed
during one XMM-Newton observation a highly absorbed power-law component and a completely
disentangled soft component. The soft component was modelled by thermal plasma emission 
\citep{2005A&A...438..211H}.
Using a thermal plasma emission component in combination with the power-law (PL+MEKAL) yields
an improved fit for \xtej\ (Table~\ref{tab-spectra}), but a temperature of $>1.3$ keV much higher than 
typical values found for other sources (0.15 keV for RX\,J0103.6$-$7201). This is inconsistent
with a very soft emission component present below $\sim$1.3 keV, similar to that observed
from RX\,J0103.6$-$7201.
A relatively hot component with characteristic blackbody temperature of $\gsim$1~keV and a 
small emission area with radius $\lsim$0.5~km was found in two persistent Be pulsars 
\citep[e.g. RX\,J0146.9+6121;][]{2006A&A...455..283L} and was interpreted as emission from the hot polar 
caps of the accreting neutron star. Adding a blackbody component to the modelling of \xtej\ (PL+BB) 
we obtain a lower temperature around 0.25 keV (Table~\ref{tab-spectra}) and a larger emission area 
(radius 30.8 km) which is inconsistent with the surface of the neutron star.
The inferred radius might indicate emission from the inner accretion disk and therefore, we
added a multi-temperature blackbody model (diskbb in XSPEC) to the power-law model 
(PL+DB in Table~\ref{tab-spectra}). Again, the fit is acceptable with a temperature of the 
inner disk of 0.36 keV (Table~\ref{tab-spectra}) and an inner disk radius of $\sim$18.7 km 
(assuming an average disk inclination angle cos\,$\theta$ of 0.5). The disk-blackbody component 
contributes 3.0\% to the total luminosity in the 0.2$-$10.0 keV band.
The EPIC spectra together with the best fit PL+DB model are shown in Fig~\ref{fig-spectra}.
In Table~\ref{tab-spectra} we summarize the characteristic model parameters and give observed 
fluxes and luminosities. 

\begin{table*}
\caption[]{Spectral fit results.}
\begin{center}
\begin{tabular}{lcccccc}
\hline\hline\noalign{\smallskip}
\multicolumn{1}{l}{Model$^{(1)}$} &
\multicolumn{1}{c}{SMC \nh} &
\multicolumn{1}{c}{$\gamma$} &
\multicolumn{1}{c}{kT} &
\multicolumn{1}{c}{Flux$^{(2)}$} &
\multicolumn{1}{c}{L$_{\rm x}^{(3)}$} &
\multicolumn{1}{c}{$\chi^2_{\rm r}$/dof} \\
\multicolumn{1}{c}{} &
\multicolumn{1}{c}{[\oexpo{20}cm$^{-2}$]} &
\multicolumn{1}{c}{} &
\multicolumn{1}{c}{[keV]} &
\multicolumn{1}{c}{erg cm$^{-2}$ s$^{-1}$} &
\multicolumn{1}{c}{erg s$^{-1}$} &
\multicolumn{1}{c}{} \\

\noalign{\smallskip}\hline\noalign{\smallskip}
 PL       & 2.44$\pm$0.85  & 0.54$\pm$0.02 & --                   & 4.55\expo{-11} & 1.98\expo{37} & 1.21/830 \\
 PL+BB    &  7.1$\pm$1.6   & 0.42$\pm$0.02 & 0.242$\pm$0.024      & 4.75\expo{-11} & 2.08\expo{37} & 1.02/828 \\
 PL+MEKAL &  9.4$\pm$1.5   & 0.35$\pm$0.08 & 1.7$^{+1.9}_{-0.4}$  & 4.79\expo{-11} & 2.11\expo{37} & 1.04/828 \\
 PL+DB    & 11.3$\pm$2.3   & 0.41$\pm$0.04 & 0.36$\pm$0.06        & 4.77\expo{-11} & 2.10\expo{37} & 1.02/828 \\
\noalign{\smallskip}\hline
\end{tabular}
\end{center}
$^{(1)}$ For definition of spectral models see text. $^{(2)}$ Observed 0.2-10.0 keV flux.
$^{(3)}$ Source intrinsic X-ray luminosity in the 0.2-10.0 keV band (corrected for absorption)
for a distance to the SMC of 60 kpc \citep{2005MNRAS.357..304H}.
\label{tab-spectra}
\end{table*}

\section{Discussion}

After its discovery with RXTE in April 2003, the 6.85 s pulsar \xtej\ was detected in X-rays for the 
first time with imaging instruments which allowed us to better localize the source and identify the 
optical counterpart. During the \xmm\ observation in October 2006 the source was seen in outburst 
with a luminosity of $\sim$2\ergs{37} in the 0.2$-$10.0 keV band. During the observation the source 
strongly varied in intensity with changes by a factor of two within ten minutes. Pulse profiles obtained 
by folding the data from the EPIC-PN detector in different energy bands with 30 phase bins 
(corresponding to 228.5 ms per phase bin) are highly structured. The fast intensity variations
indicate that the bulk of the emission originates close to the neutron star surface and suggests 
a complicated emission geometry with narrow beams contributing to the X-ray emission at various 
neutron star spin phases and with energy dependent amplitudes.

Deviations from a power-law model in the phase averaged EPIC spectra suggest an additional emission component
mainly contributing at energies below $\sim$1.5~keV. This component is harder than the soft component seen
in other Be-HMXBs in the SMC, which can be modeled by thermal plasma emission with kT$\sim$0.15 keV
\citep[see e.g. RX\,J0103.6$-$7201;][]{2005A&A...438..211H}. On the other hand the excess emission
in the spectra of \xtej\ is softer than the hot blackbody component seen in \xmm\ spectra of 
two persistent Be-HMXBs in the Milky Way \citep{2006A&A...455..283L,2007A&A...474..137L} which
is interpreted as emission from hot polar caps of the neutron star. When modelled as blackbody
emission, the large inferred emission area (with radius $\sim$30 km) for \xtej\ is incompatible with
an origin on the neutron star surface.  

In their work on the origin of the soft excess in X-ray pulsars, \citet{2004ApJ...614..881H}
conclude that for luminous X-ray pulsars ($\gsim$\oergs{38}) the soft excess
can only be explained by reprocessing of hard X-rays from the neutron star by 
optically thick material. At intermediate luminosity ($\sim$\oergs{37}) also other processes
such as emission from photo-ionized or collisionally heated gas can contribute.
The spectra of \xtej\ do not show significant line emission suggesting that in this source
the excess emission is mainly caused by reprocessing in optically thick material, most likely 
located near the inner edge of an accretion disk. The relatively low contribution ($\sim$3\%) 
to the total luminosity and the estimate for the inner disk radius of $\sim$19~km inferred 
from the disk-blackbody model are consistent with such a model. 

The luminosity of the soft component seen from RX\,J0103.6$-$7201 is strongly correlated 
with the total source luminosity over at least a factor of ten variation in source intensity.
At a maximum of $\sim$6.4\ergs{36} RX\,J0103.6$-$7201 is still a factor of $\sim$3 below
the luminosity of \xtej\ during its October 2006 outburst. This is consistent with the 
conclusions of \citet{2004ApJ...614..881H}, that for lower luminosities emission from diffuse, 
optically thin gas dominates. Which of the emission components becomes visible in the X-ray spectra
of Be-HMXBs certainly depends not only on source luminosity but also on geometric effects.

\begin{acknowledgements}
The XMM-Newton project is supported by the Bundesministerium f\"ur Wirtschaft und 
Technologie/Deutsches Zentrum f\"ur Luft- und Raumfahrt (BMWI/DLR, FKZ 50 OX 0001)
and the Max-Planck Society. 
\end{acknowledgements}

\bibliographystyle{aa}
\bibliography{general,myrefereed,myunrefereed,mcs,hmxb,ism}

\begin{thebibliography}{24}
\expandafter\ifx\csname natexlab\endcsname\relax\def\natexlab#1{#1}\fi

\bibitem[{{Coe} {et~al.}(2005){Coe}, {Edge}, {Galache}, \&
  {McBride}}]{2005MNRAS.356..502C}
{Coe}, M.~J., {Edge}, W.~R.~T., {Galache}, J.~L., \& {McBride}, V.~A. 2005,
  \mnras, 356, 502

\bibitem[{{Corbet} {et~al.}(2003){Corbet}, {Markwardt}, {Marshall}, {Coe},
  {Edge}, \& {Laycock}}]{2003ATel..163....1C}
{Corbet}, R.~H.~D., {Markwardt}, C.~B., {Marshall}, F.~E., {et~al.} 2003, The
  Astronomer's Telegram, 163, 1

\bibitem[{{Haberl} {et~al.}(2000){Haberl}, {Filipovi{\'c}}, {Pietsch}, \&
  {Kahabka}}]{2000A&AS..142...41H}
{Haberl}, F., {Filipovi{\'c}}, M.~D., {Pietsch}, W., \& {Kahabka}, P. 2000,
  \aaps, 142, 41

\bibitem[{{Haberl} \& {Pietsch}(2004)}]{2004A&A...414..667H}
{Haberl}, F. \& {Pietsch}, W. 2004, \aap, 414, 667

\bibitem[{{Haberl} \& {Pietsch}(2005)}]{2005A&A...438..211H}
{Haberl}, F. \& {Pietsch}, W. 2005, \aap, 438, 211

\bibitem[{{Haberl} {et~al.}(2007){Haberl}, {Pietsch}, \&
  {Kahabka}}]{2007ATel.1095....1H}
{Haberl}, F., {Pietsch}, W., \& {Kahabka}, P. 2007, The Astronomer's Telegram,
  1095, 1

\bibitem[{{Hickox} {et~al.}(2004){Hickox}, {Narayan}, \&
  {Kallman}}]{2004ApJ...614..881H}
{Hickox}, R.~C., {Narayan}, R., \& {Kallman}, T.~R. 2004, \apj, 614, 881

\bibitem[{{Hilditch} {et~al.}(2005){Hilditch}, {Howarth}, \&
  {Harries}}]{2005MNRAS.357..304H}
{Hilditch}, R.~W., {Howarth}, I.~D., \& {Harries}, T.~J. 2005, \mnras, 357, 304

\bibitem[{{Jansen} {et~al.}(2001){Jansen}, {Lumb}, {Altieri}, {Clavel}, {Ehle},
  {Erd}, {Gabriel}, {Guainazzi}, {Gondoin}, {Much}, {Munoz}, {Santos},
  {Schartel}, {Texier}, \& {Vacanti}}]{2001A&A...365L...1J}
{Jansen}, F., {Lumb}, D., {Altieri}, B., {et~al.} 2001, \aap, 365, L1

\bibitem[{{La Palombara} \& {Mereghetti}(2006)}]{2006A&A...455..283L}
{La Palombara}, N. \& {Mereghetti}, S. 2006, \aap, 455, 283

\bibitem[{{La Palombara} \& {Mereghetti}(2007)}]{2007A&A...474..137L}
{La Palombara}, N. \& {Mereghetti}, S. 2007, \aap, 474, 137

\bibitem[{{Laycock} {et~al.}(2005){Laycock}, {Corbet}, {Coe}, {Marshall},
  {Markwardt}, \& {Lochner}}]{2005ApJS..161...96L}
{Laycock}, S., {Corbet}, R.~H.~D., {Coe}, M.~J., {et~al.} 2005, \apjs, 161, 96

\bibitem[{{Massey}(2002)}]{2002ApJS..141...81M}
{Massey}, P. 2002, \apjs, 141, 81

\bibitem[{{McGowan} {et~al.}(2007){McGowan}, {Coe}, {Schurch}, {Corbet},
  {Galache}, \& {Udalski}}]{2007arXiv0711.4010M}
{McGowan}, K.~E., {Coe}, M.~J., {Schurch}, M.~P.~E., {et~al.} 2007, ArXiv
  e-prints, 711

\bibitem[{{Negueruela}(1998)}]{1998A&A...338..505N}
{Negueruela}, I. 1998, \aap, 338, 505

\bibitem[{{Okazaki} \& {Negueruela}(2001)}]{2001A&A...377..161O}
{Okazaki}, A.~T. \& {Negueruela}, I. 2001, \aap, 377, 161

\bibitem[{{Russell} \& {Dopita}(1992)}]{1992ApJ...384..508R}
{Russell}, S.~C. \& {Dopita}, M.~A. 1992, \apj, 384, 508

\bibitem[{{Sasaki} {et~al.}(2000){Sasaki}, {Haberl}, \&
  {Pietsch}}]{2000A&AS..147...75S}
{Sasaki}, M., {Haberl}, F., \& {Pietsch}, W. 2000, \aaps, 147, 75

\bibitem[{{Schmidtke} \& {Cowley}(2007)}]{2007ATel.1181....1S}
{Schmidtke}, P.~C. \& {Cowley}, A.~P. 2007, The Astronomer's Telegram, 1181, 1

\bibitem[{{Str{\"u}der} {et~al.}(2001){Str{\"u}der}, {Briel}, {Dennerl},
  {Hartmann}, {Kendziorra}, {Meidinger}, {Pfeffermann}, {Reppin}, {Aschenbach},
  {Bornemann}, {Br{\"a}uninger}, {Burkert}, {Elender}, {Freyberg}, {Haberl},
  {Hartner}, {Heuschmann}, {Hippmann}, {Kastelic}, {Kemmer}, {Kettenring},
  {Kink}, {Krause}, {M{\"u}ller}, {Oppitz}, {Pietsch}, {Popp}, {Predehl},
  {Read}, {Stephan}, {St{\"o}tter}, {Tr{\"u}mper}, {Holl}, {Kemmer}, {Soltau},
  {St{\"o}tter}, {Weber}, {Weichert}, {von Zanthier}, {Carathanassis}, {Lutz},
  {Richter}, {Solc}, {B{\"o}ttcher}, {Kuster}, {Staubert}, {Abbey}, {Holland},
  {Turner}, {Balasini}, {Bignami}, {La Palombara}, {Villa}, {Buttler},
  {Gianini}, {Lain{\'e}}, {Lumb}, \& {Dhez}}]{2001A&A...365L..18S}
{Str{\"u}der}, L., {Briel}, U., {Dennerl}, K., {et~al.} 2001, \aap, 365, L18

\bibitem[{{Turner} {et~al.}(2001){Turner}, {Abbey}, {Arnaud}, {Balasini},
  {Barbera}, {Belsole}, {Bennie}, {Bernard}, {Bignami}, {Boer}, {Briel},
  {Butler}, {Cara}, {Chabaud}, {Cole}, {Collura}, {Conte}, {Cros}, {Denby},
  {Dhez}, {Di Coco}, {Dowson}, {Ferrando}, {Ghizzardi}, {Gianotti}, {Goodall},
  {Gretton}, {Griffiths}, {Hainaut}, {Hochedez}, {Holland}, {Jourdain},
  {Kendziorra}, {Lagostina}, {Laine}, {La Palombara}, {Lortholary}, {Lumb},
  {Marty}, {Molendi}, {Pigot}, {Poindron}, {Pounds}, {Reeves}, {Reppin},
  {Rothenflug}, {Salvetat}, {Sauvageot}, {Schmitt}, {Sembay}, {Short},
  {Spragg}, {Stephen}, {Str{\"u}der}, {Tiengo}, {Trifoglio}, {Tr{\"u}mper},
  {Vercellone}, {Vigroux}, {Villa}, {Ward}, {Whitehead}, \&
  {Zonca}}]{2001A&A...365L..27T}
{Turner}, M. J.~L., {Abbey}, A., {Arnaud}, M., {et~al.} 2001, \aap, 365, L27

\bibitem[{{Udalski} {et~al.}(1998){Udalski}, {Szymanski}, {Kubiak},
  {Pietrzynski}, {Wozniak}, \& {Zebrun}}]{1998AcA....48..147U}
{Udalski}, A., {Szymanski}, M., {Kubiak}, M., {et~al.} 1998, Acta Astronomica,
  48, 147

\bibitem[{{Wilms} {et~al.}(2000){Wilms}, {Allen}, \&
  {McCray}}]{2000ApJ...542..914W}
{Wilms}, J., {Allen}, A., \& {McCray}, R. 2000, \apj, 542, 914

\bibitem[{{Zaritsky} {et~al.}(2002){Zaritsky}, {Harris}, {Thompson}, {Grebel},
  \& {Massey}}]{2002AJ....123..855Z}
{Zaritsky}, D., {Harris}, J., {Thompson}, I.~B., {Grebel}, E.~K., \& {Massey},
  P. 2002, \aj, 123, 855

\end{thebibliography}

\end{document}